\documentclass[floats,floatfix,showpacs,amssymb,prd,twocolumn,superscriptaddress,nofootinbib]{revtex4-1}

\usepackage{amssymb,amsmath,verbatim,mathtools,needspace,enumitem,etoolbox,graphicx,physics,microtype,afterpage,bm}

\usepackage{subcaption}
\usepackage{ragged2e}
\DeclareCaptionJustification{justified}{\justifying}
\usepackage[dvipsnames, usenames]{xcolor}
\usepackage[normalem]{ulem}

\definecolor{linkcolor}{rgb}{0.0,0.3,0.5}
\usepackage[unicode, colorlinks=true, linkcolor=linkcolor, citecolor=linkcolor, filecolor=linkcolor,urlcolor=linkcolor, pdfusetitle]{hyperref}
\usepackage[all]{hypcap}
\usepackage[T1]{fontenc}
\usepackage[utf8]{inputenc}
\usepackage{tabularx}
\usepackage{float}
\allowdisplaybreaks
\graphicspath{{./figure/}}

\newcommand{\llp}{\left [}
\newcommand{\rrp}{\right ]}

\newcommand{\lp}{\left (}
\newcommand{\rp}{\right )}

\newcommand{\ita}{\text{\tiny ita}}
\newcommand{\ta}{\text{\tiny ta}}

\def\PBH{\text{\tiny PBH}}

\def\tt{\text{\tiny t}}
\newcommand{\be}{\begin{equation}\begin{aligned}}
\newcommand{\ee}{\end{aligned}\end{equation}}

\newcommand{\bbe}{\begin{align}}
\newcommand{\eee}{\end{align}}

\newcommand{\bea}{\begin{eqnarray}}
\newcommand{\eea}{\end{eqnarray}}

\def\beq{\begin{equation}}
\def\eeq{\end{equation}}

\def\d{{\rm d}}

\def\beqa{\begin{eqnarray}}

	\def\eeqa{\end{eqnarray}}
	
\def\lsim{\mathrel{\rlap{\lower4pt\hbox{\hskip0.5pt$\sim$}}
		\raise1pt\hbox{$<$}}}   %less than or approx. symbol
\def\gsim{\mathrel{\rlap{\lower4pt\hbox{\hskip0.5pt$\sim$}}
		\raise1pt\hbox{$>$}}}   %greater than or approx. symbol

\def\d{{\rm d}}

\def\d{{\rm d}}

\def\PBH{\text{\tiny \rm PBH}}

\def\eeqa{\end{eqnarray}}

\def\bq{\begin{quote}}
\def\eq{\end{quote}}

 at 10truept

\def\eeqa{\end{eqnarray}}
\def\lsim{\mathrel{\rlap{\lower4pt\hbox{\hskip0.5pt$\sim$}}
 \raise1pt\hbox{$<$}}}   %less than or approx. symbol
\def\gsim{\mathrel{\rlap{\lower4pt\hbox{\hskip0.5pt$\sim$}}
 \raise1pt\hbox{$>$}}}   %greater than or approx. symbol

\def\apj{\rm{ApJ}}

\definecolor{rb4}{HTML}{27408B}

\begin{document}

\title{Primordial Black Holes in Matter-Dominated Eras: the Role of Accretion}

\author{Valerio~De~Luca}
\email{Valerio.DeLuca@unige.ch}
\affiliation{D\'epartement de Physique Th\'eorique and Centre for Astroparticle Physics (CAP), Universit\'e de Gen\`eve, 24 quai E. Ansermet, CH-1211 Geneva, Switzerland}

\author{Gabriele~Franciolini}
\email{gabriele.franciolini@uniroma1.it}
\affiliation{Dipartimento di Fisica, Sapienza Università di Roma, Piazzale Aldo Moro 5, 00185, Roma, Italy}

\author{Alex~Kehagias}
\email{Kehagias@central.ntua.gr}
\affiliation{Physics Division, National Technical University of Athens, Zografou, Athens, 15780, Greece}

\author{Paolo~Pani}
\email{paolo.pani@uniroma1.it}
\affiliation{Dipartimento di Fisica, Sapienza Università di Roma, Piazzale Aldo Moro 5, 00185, Roma, Italy}
\affiliation{INFN, Sezione di Roma, Piazzale Aldo Moro 2, 00185, Roma, Italy}

\author{Antonio~Riotto}
\email{Antonio.Riotto@unige.ch}
\affiliation{D\'epartement de Physique Th\'eorique and Centre for Astroparticle Physics (CAP), Universit\'e de Gen\`eve, 24 quai E. Ansermet, CH-1211 Geneva, Switzerland}

\date{\today}

\begin{abstract} \noindent
We consider the role of secondary infall and accretion onto an initially overdense perturbation in  matter-dominated eras, like the one which is likely to follow the end of inflation. We show that primordial black holes may form through post-collapse accretion, namely the accretion onto an initial overdensity whose collapse has not given rise to a  primordial black hole. Accretion may be also responsible for the growth of the primordial black hole masses  by orders of magnitude till the end of the matter-dominated era. 
\end{abstract}

\maketitle

\paragraph*{\it 1. Introduction.} 
\noindent
Primordial Black Holes (PBHs) have attracted lot of interest \cite{Sasaki:2018dmp,Carr:2020gox,Green:2020jor} since the
 detection of gravitational waves generated by the mergers of two black holes \cite{LIGOScientific:2016aoc,LIGOScientific:2018mvr, LIGOScientific:2020ibl,LIGOScientific:2021djp} as they could be contributing to a fraction of the events observed by the LIGO/Virgo/KAGRA collaboration \cite{Hutsi:2020sol,DeLuca:2021wjr,Franciolini:2021tla}. 
 
 PBHs may be formed due to the collapse of large overdensities in the early stages of the evolution of the universe. 
If the collapse takes place during a radiation-dominated phase, PBHs are generated only if the initial amplitude of the density perturbation is beyond a large threshold (for recent studies, see Refs.~\cite{Musco:2018rwt,Musco:2020jjb})
due to the counteracting action of the radiation pressure. 

Such an impediment is basically absent in the case in which PBHs are formed in a pressureless environment, that is in a matter-dominated phase~\cite{Harada:2016mhb, Harada:2017fjm, Kokubu:2018fxy}. The latter can take place, for instance, after inflation when the scalar field (the inflaton), whose vacuum energy during inflation was responsible for the accelerated phase, effectively behaves
like a pressureless fluid when oscillating around the minimum of its potential. At this stage the universe is filled by the classical inflaton field and by its quanta whose fluctuations are gradually re-entering the horizon. They may give rise, upon collapse, to bounded structures~\cite{Niemeyer:2019gab,Musoke:2019ima,Eggemeier:2020zeg,Chen:2020cef,Eggemeier:2021smj} and eventually to PBHs~\cite{deJong:2021bbo, Padilla:2021zgm}. This early matter-dominated phase ends when the inflaton decays and reheats the universe, starting the radiation-dominated phase. Another possibility of early matter-dominated phase is a period dominated by the oscillations of moduli fields, which are ubiquitous in string theory, and have only gravitational interactions (see for instance Ref.~\cite{Coughlan:1983ci}). 
Of course, one might consider as well the standard late matter-dominated phase,  even though it  corresponds to  PBH masses which are too large to be of any observational interest.

In this paper, we will be interested in the role of accretion in the formation and mass growth of PBHs during a matter-dominated period, a subject which has been poorly investigated, to the best of our knowledge, in the literature (see, however, Ref. \cite{deJong:2021bbo} for a recent numerical study). In particular, we will study under which conditions PBHs may form thanks to the accretion onto an initial overdensity which has not collapsed into a PBH, but in a dispersed cloud or a bosonic solitonic star, a mechanism dubbed post-collapse accretion. Accretion may also be relevant in increasing the mass of an initial PBH (whatever its origin is, direct production or through post-collapse accretion) during the matter-dominated era and we will also devote our attention to this possibility, distinguishing  the cases in which the Compton wavelength of the surrounding particles is smaller or larger than the PBH horizon.

Our considerations start with the presence, in a matter-dominated phase, of an initial density perturbation. After its collapse,  bound shells of the surrounding material continue to turn around and fall in, a process dubbed secondary infall~\cite{Bertschinger:1985pd}. If spherical symmetry is assumed, the secondary infall is described by self-similar solutions with the turn-around radius of the shells increasing with time. Even though  infinitesimal deviations from spherical symmetry are magnified during a uniform collapse~\cite{1965ApJ...142.1431L}, 
 the estimates obtained assuming spherical symmetry can remain valid
if the role of angular momentum proves to be minor, a condition we will return to later on.

\begin{figure*}[th]
	\includegraphics[width=0.55 \linewidth]{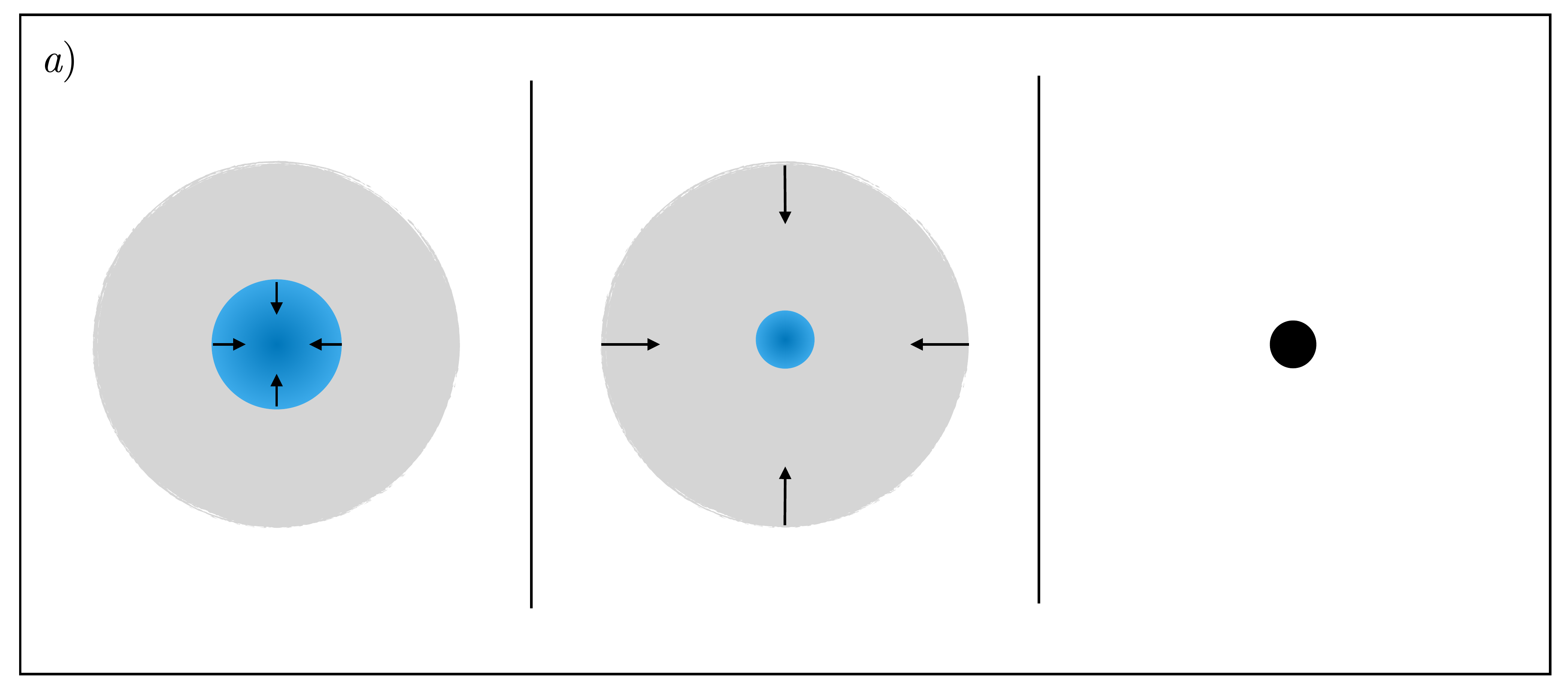}
	\includegraphics[width=0.36 \linewidth]{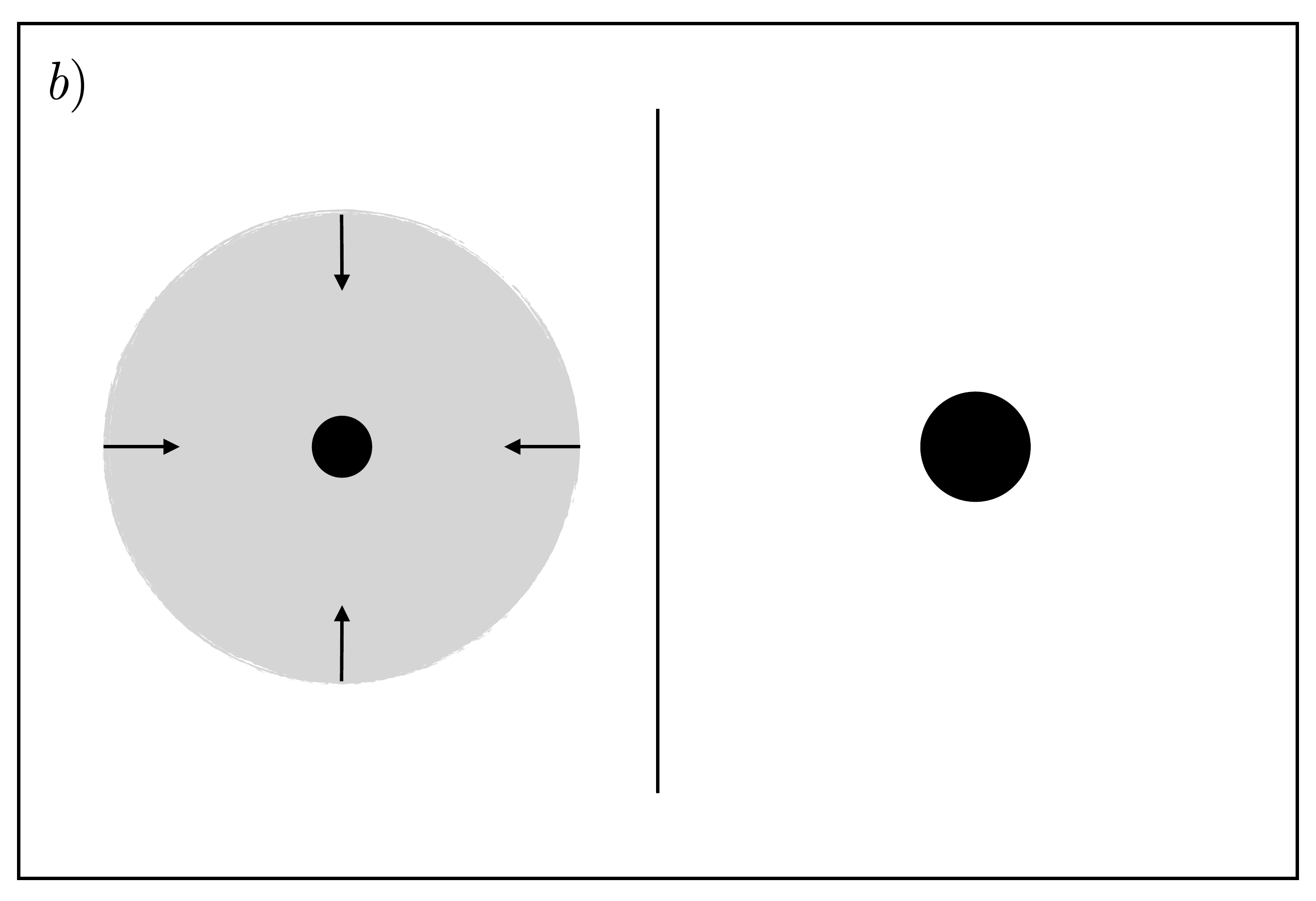}
	\caption{\it a) Pictorial representation of the formation of a PBH through the post-collapse accretion phase; b) Pictorial representation of the mass growth of a PBH via accretion for the surrounding material.
	}
	\label{fig:1}
\end{figure*}

The impact of secondary infall and accretion from the surrounding non-relativistic gas onto the collapsed structures depends on their nature.
If the structures are not PBHs, they may, in principle, rapidly accrete the surrounding material and give rise to PBHs. We will show that, under the  spherical symmetry assumption, PBHs may form by post-collapse accretion onto pre-existing matter overdensities like self-gravitating clouds or bounded clumps, of which a fascinating possibility is provided by inflaton stars~\cite{Niemeyer:2019gab,Musoke:2019ima,Eggemeier:2020zeg,Chen:2020cef,Eggemeier:2021smj}. We will also show that PBHs may efficiently accrete and increase their initial mass by various orders of magnitude during the matter-dominated era, 
provided the geometry of the accretion flow remains nearly spherical. 
A pictorial representation of the two scenarios is provided in Fig.~\ref{fig:1}.

\vskip 0.3cm
\noindent
\paragraph*{\it 2. PBHs formation through post-collapse accretion.} Consider a sphere of radius $R_i$ with an uniform overdensity $\delta\rho/\rho=\delta_i\ll 1$ at the cosmic time $t_i$ during the matter-dominated phase,  such that the total energy density reads
\be
\rho(t_i,r)=\frac{1}{6\pi G t_i^2}\left\{
\begin{array}{cc}
1+\delta_i & r<R_i\\
1 & r>R_i
\end{array}\right.\,,
\ee
where $G$ is Newton's constant. The radius $R_i$ may be identified with the size of a peak of amplitude $\delta_i$ which has re-entered the horizon at a time before $t_i = 2/3H_i$. As the density contrast grows, the material inside $R_i$ is decelerated, it ceases to expand, and it turns around to recollapse at the time $t_\ita=(3\pi/4)\delta_i^{-3/2} t_i$, when the maximum radius of the comoving shell initially at $R_i$ is $r_\ita= R_i\delta_i^{-1}$. After the first mass shell has turned around, the matter in the overdense region collapses, reaching a putative infinite density at a time $\approx 2t_\ita$. 

After the collapse, the motion of the surrounding fluid depends on the leftover structure and the nature of the fluid. We will be first concerned, in section 2.1,
with the appearance of a dispersed cloud of collisionless fluid   generating a strong attractive gravitational potential. This case may describe both the dynamics during the standard matter-dominated phase, as well as the dynamics during an early matter-dominated phase when the universe is filled by a scalar field (inflaton or modulus), but the size of the collapsed cloud is such that quantum effects play no role and no bosonic soliton is formed. 
Subsequently, in section 2.2, we will analyse  accretion onto overdense  solitonic stars in the case in which, for instance,  the early matter-dominated phase is the one immediately after inflation~\cite{Niemeyer:2019gab,Musoke:2019ima,Eggemeier:2020zeg,Chen:2020cef,Eggemeier:2021smj}. In both cases, we assume that no PBHs have formed from the collapse of the initial overdensity. In other words, we assume that the hoop conjecture~\cite{Misner:1973prb}, indicating the formation of a black hole when a circular hoop with a specific critical circumference may be placed around the object and rotated about its diameter whose size is smaller than the Schwarzschild radius, is not satisfied.~\footnote{ While the hoop conjecture has been formulated in asymptotically flat spacetime, it should equally apply in an expanding universe as long as the cosmological horizon is much larger than the Schwarzschild
radius, i.e. $1/H\gg 2 G M$, where $H$ is the background Hubble rate.}

\vskip 0.3cm
\noindent
\paragraph*{\it 2.1} 
Let us start with the case where there is a cloud onto which the surrounding matter may accrete. As mentioned before, this case applies to both the standard and the early matter-dominated phase (in which no solitonic star has been formed).  The mass around the structure will start rapidly accumulating through the process of secondary infall \cite{Bertschinger:1985pd}. Particles turning around at a cosmic time $t= t_\ta\gg t_\ita$ reach the maximum radius $r_\ta (t_\ta)\propto t_\ta^{8/9}$. The mass contained within a radius $r \ll r_\ta(t_\ta)$ is found to be constant in time since it is dominated by shells that have turned around at earlier times ($t< t_\ta$). Following Refs.~\cite{Gunn:1972sv, Berezinsky:2013fxa} and defining the adimensional radius $\lambda=r/r_\ta$, the secondary infall results in a density profile 
\be
\rho(r \ll r_\ta)\simeq 3\overline\rho\, \lambda^{-9/4},
\ee
where $\overline\rho=3H^2/8\pi G$ is the background energy density and $H$ is the corresponding Hubble rate. The mass contained within a radius $r \ll r_\ta$ is therefore
\be
M(<r)\simeq 16\pi \overline\rho r_\ta^{9/4}r^{3/4},
\ee
which, as remarked above, does not depend upon time.

Satisfying the hoop conjecture to form a PBH from the secondary infall requires
\be
r\lsim 2 G M(<r).
\ee
Using the relation for the turn-around radius $r_\text{\tiny ta}\simeq r_\text{\tiny ita} (t/t_\text{\tiny ita})^{8/9}$, we find that a PBH can form within a radius satisfying the upper bound
\be
r\lsim 20 H_i^8 R_i^9 \delta_i^3.
\ee
This corresponds to a maximum PBH mass of
\be
M_\PBH^\text{\tiny max} \simeq 20\, \delta_i^3 M_{\text{\tiny H},i},
\ee
where we have used that $R_i \lesssim H_i^{-1}$ and expressed the initial Hubble radius in terms of the corresponding horizon mass $M_{\text{\tiny H},i} = 1/ (2 G H_i)$.

This result suggests that the formation of PBHs through post-collapse accretion of a non-relativistic collisionless gas 
is indeed possible and can occur within a Hubble time. We remark though that small deviations from sphericity are amplified in the infall~\cite{1965ApJ...142.1431L}, resulting in a collapse taking place along a pancake or a line rather than a point (see also Ref.~\cite{Saenz:1978mc}).
This implies the generation of angular momentum which would tend to halt the PBH formation. 
Furthermore, we stress again that we have neglected the possible presence of self-interactions of the accreting particles (e.g. the inflaton quanta) when dealing with an early matter-dominated phase.

\vskip 0.3cm
\noindent
\paragraph*{\it 2.2} 

The second scenario arises when one resolves the quantum nature of the scalar  field responsible for the pressureless phase, e.g. the inflaton field during the reheating phase. In such a case  the Compton wavelength of the inflaton $\sim 1/m$ ($m$ being the mass of the inflaton) is comparable to, or larger than, the typical size of the collapsed object.
In such a  case, the formation of PBHs may occur from accretion onto inflaton stars, which are localized lumps of boson energy density held together by the competing forces of gravity, self-interactions and gradient energy, and 
% may
{\rm could}
form during the post-inflation early matter-dominated phase and before reheating~\cite{Niemeyer:2019gab,Musoke:2019ima,Eggemeier:2020zeg,Chen:2020cef,Eggemeier:2021smj}. In the following, we will assume that the initial overdensity produced by the scalar field inhomogeneity is not so large to give rise directly to a PBH, but instead it forms a solitonic core. 

For bosons without self-interactions, this process may happen through gravitational cooling~\cite{Seidel:1993zk,PhysRevD.42.384,Alcubierre:2003sx}, in which early miniclusters of the surrounding scalar field formed due to gravitational interactions evolve by expelling probability density to infinity since all eigenstates other than the solitonic ground state are unstable (see instead Refs.~\cite{1986SvAL...12..305T,Kolb:1993hw,Kolb:1993zz,Eggemeier:2019jsu} for solitons generated from axion interactions). This process occurs 
within a characteristic timescale of the order of the relaxation time~\cite{1987gady.book.....B, Levkov:2018kau}, defined as
\be\label{eqtstar}
t_* = 0.08 \frac{v_*^3}{G^2 m_\text{\tiny eff} \rho \Lambda_*} \simeq 2.5 \cdot 10^{-3} \frac{m^3 v_*^6}{G^2 \rho^2 \Lambda_*},
\ee
in which the minicluster behaves like a halo of quasiparticles with effective mass $m_\text{\tiny eff} \simeq \rho (\lambda/2)^3 \sim \pi^3\rho/m^3 v_*^3$~\cite{Hui:2016ltb}, in terms of the local density $\rho$, de Broglie wavelength $\lambda = 2 \pi/m v_*$ and soliton velocity $v_*$.
The properties of the surrounding minicluster, i.e. 
its mass $M_h$, radius $r_h$ and virial velocity $v_\text{\tiny vir} = \sqrt{G M_h/r_h}$, 
enter in Eq.~\eqref{eqtstar} through the parameter 
$\Lambda_* \equiv \log \lp m v_* r_h \rp$.

After nucleation, the presence of the surrounding inflaton halo induces a phase of matter accretion onto the stars, with a mass growth rate found numerically as~\cite{Levkov:2018kau}
\be
\label{accretion1}
M_* (t) \simeq M_i \lp \frac{t}{t_i} \rp^{1/2},
\ee
in terms of the initial soliton star mass $M_i$ at time $t_i \sim t_*(M_i) \sim 0.2/ ( m^3 G^2 M_i^2)$.
We now show that
this behaviour of the soliton mass can be explained analytically following the dynamics of the core collapse model~\cite{1987gady.book.....B}. Indeed, the outer parts of the minicluster expand by expelling probability density to infinity, and at the same time, the centre contracts with a dramatic growth in the central density. 
At radii much larger than the core size $r_* (t)$, the density profile evolves self-similarly, with a solution of the form
\be
\rho (r,t) \equiv \rho_* (t) \tilde \rho (r/r_*).
\ee
The energy density is found to be nearly independent of time, as shown also in the numerical simulation of Ref.~\cite{Chen:2020cef}, and one thus gets
\be
\frac{\partial \rho}{\partial t} = \frac{\d \rho_*}{\d t}\tilde \rho - \rho_* \frac{\d \tilde \rho}{\d (r/r_*)} \frac{r}{r_*^2} \frac{\d r_*}{\d t} = 0,
\ee
which can be rearranged as
\be
\frac{r_*}{\rho_*} \frac{\d \rho_*}{\d t} \frac{\d t}{\d r_*} = \frac{\d \tilde \rho}{\d (r/r_*)} \frac{(r/r_*)}{\tilde \rho}.
\ee
Given that the left-hand side is independent of the radial coordinate and the right-hand side is independent of the time coordinate, they must be equal to a constant, which we denote as $-\alpha$, such that the solution is described by the power laws
\begin{align}
\rho_* (t) &\propto r_*^{-\alpha}(t), \nonumber \\
\tilde \rho (r/r_*)&\propto(r/r_*)^{-\alpha}.
\end{align}
The slope parameter can be determined by realising that, during the core collapse, the higher modes of the inflaton star decay and for all of them (including the zero mode)  the core mass relation $M_*(t) \equiv \rho_* (t) r_*^3 (t) \sim 4/(G m^2 r_* (t))$ is valid due to the balance between gravity and quantum pressure~\cite{Hui:2016ltb}. In this way one gets $\alpha \simeq 4$.
Taking into account that $t_*\propto v_*^6/\rho_*^2 \propto r_*^{2}$, where $v_*\simeq G M_* m\propto 1/r_*$ and $\rho_*\propto M_*^4\propto r_*^{-4}$~\cite{Hui:2016ltb}, during the core collapse the core radius contracts as 
\be
\frac{\dot r_*}{r_*} \simeq -\frac{1}{t_*}\propto -r_*^{-2} \quad {\rm or} \quad r_*(t)\propto r_* (t_i)\lp\frac{3t_i-2t}{t_i}\rp^{1/2},
\ee
valid up to the virialization time scale $t_\text{\tiny vir} = t_*(v_* = v_\text{\tiny vir})$~\cite{Levkov:2018kau}, when 
 it becomes comparable to the characteristic size of the granular structures contained in the 
surrounding halos of the order of the de Broglie wavelength $\lambda \sim r_*(t_\text{\tiny vir})\sim (m v_\text{\tiny vir})^{-1}$.
Therefore, the corresponding soliton mass grows with time 
reproducing the behaviour in Eq.~\eqref{accretion1} by expanding for times close to  $t_i$.

At the virialization time, the surrounding halos will reach virial equilibrium with the soliton, causing the mass growth to slow down and continue at a drastically reduced rate as~\cite{Eggemeier:2019jsu,Chen:2020cef}
\be
\label{accretion2}
M_* (t) \simeq M_*(t_\text{\tiny vir}) \lp \frac{t}{t_\text{\tiny vir}}\rp^{1/8},
\ee
where the time dependence slope has been deduced by assuming that the power law behaviour continues to hold after the virialization time and evaluating Eq.~\eqref{eqtstar} with $v_* = v_\text{\tiny vir}$~\cite{Eggemeier:2019jsu}.
The accretion onto the inflaton star may induce its collapse into a PBH.\footnote{See, for example, Ref.~\cite{Helfer:2016ljl} for a simulation of PBH formation from axion stars due to accretion during a late matter-dominated era.}
This happens when the Kaup limit is reached~\cite{Kaup:1968zz,PhysRev.187.1767,Helfer:2016ljl},
that is when the Compton wavelength becomes smaller than the Schwarzschild radius, or 
\be
M_*(t)\gtrsim M_\text{\tiny Kaup} \equiv \frac{0.6}{G m}.
\ee
The collapse into a PBH may happen in both the accretion phases before or after the virialization time. In particular, for the first regime in Eq.~\eqref{accretion1} one gets the condition 
\be
\frac{t}{t_i} \gtrsim \lp \frac{0.2}{G M_{i} m}\rp^2 \qquad \text{for} \qquad t \lesssim t_\text{\tiny vir},
\ee
while, for the second regime in Eq.~\eqref{accretion2}, one obtains
\be
\frac{t}{t_\text{\tiny vir}} \gtrsim \lp \frac{0.2}{v_\text{\tiny vir}}\rp^8 \qquad \text{for} \qquad t \gtrsim t_\text{\tiny vir}.
\ee
One can appreciate how both criteria depend on the initial conditions of soliton formation and characteristic halos surrounding it. Of course, PBHs  form only if reheating takes place at sufficiently late times to allow the accretion to be efficient enough. One can therefore derive an upper bound on the reheating temperature for this dynamics to take place 
\be
T_\text{\tiny RH} \lesssim G^{-3/4}  M_i^2 M_\text{\tiny Kaup}^{-5/2}
\ee
for the first regime, and 
\be
T_\text{\tiny RH} \lesssim 70 \, G^{-3/4} v_\text{\tiny vir}^{5} M_\text{\tiny Kaup}^{-1/2}
\ee
for the second regime. In units of the inflaton mass and the Planck mass $M_\text{\tiny p} = G^{-1/2}$, the above bounds become 
\begin{align}
   T_\text{\tiny RH} &\lesssim 4 \lp \frac{m M_i}{M_\text{\tiny p}^2}\rp^2 \sqrt{M_\text{\tiny p} m},\\
T_\text{\tiny RH} &\lesssim 90 \, v_\text{\tiny vir}^{5} \sqrt{M_\text{\tiny p} m}, 
\end{align}
in the two regimes, respectively. These have to be compared with the reheating temperature $T_\text{\tiny RH} \sim 0.1 \sqrt{\Gamma M_\text{\tiny p}}$, where $\Gamma= \gamma m$ is the inflaton decay rate and $\gamma \ll 1$ parametrises the coupling constant of the inflaton field to the Standard Model particles.
We conclude that spherical accretion is efficient enough for causing PBH formation  depending on the epoch at which reheating occurs, on the properties of the surrounding halos  and for overdensities heavier than 
\be
M_i \gtrsim 0.2 \gamma^{1/4} \frac{M_p^2}{m}.
\ee

\vskip 0.3cm
\noindent
\paragraph*{\it 3. PBH mass growth from accretion.} Let us suppose now that the initial overdensity $\delta_i$ has given rise to a PBH, either by direct collapse or by post-collapse accretion. We wish to show under which conditions PBH masses may grow due to accretion. We will consider both cases in which the Compton wavelength of the (bosonic) particles is smaller and larger than or comparable to the PBH horizon.

\vskip 0.3cm
\noindent
\paragraph*{\it 3.1} Let us first consider the case of accretion induced by  
a non-relativistic fluid, whose particles characteristic wavelength is much smaller than the PBH horizon. To do so, we start with the equation governing the  accretion rate of the PBH mass $M_\PBH$ \cite{Bertschinger:1985pd}
\be
\label{a}
\dot M_\PBH = \lim_{r\rightarrow 0} 4\pi r^2\rho(r) v(r).
\ee
Here $v(r)$ is the velocity of the non-relativistic particles at radius $r$ and $\rho(r)$ the corresponding energy density.
Describing again the accretion process as secondary infall onto a central black hole, one can express the self-similar solutions in terms of the dimensionless parameter $\lambda=r/r_\ta$ as~\cite{Berezinsky:2013fxa}
\begin{eqnarray}
v(r,t)&=&\frac{r_\ta}{t}V(\lambda),\nonumber\\
\rho(r)&=&\overline\rho D(\lambda).
\end{eqnarray}
At $r\ll r_\ta$, one finds
\begin{eqnarray}
V(\lambda \ll 1) &\sim &2^{-4/3} \pi \lambda^{-1/2},\nonumber\\
D(\lambda\ll 1) &\sim& 2^{ -7/3} \pi \lambda^{-3/2}.
\end{eqnarray}
Therefore, Eq.~\eqref{a} can be solved in the vicinity of the PBH to provide the change with time of the PBH mass
\be
\label{massevoaccretion}
M_\PBH (t\gsim t_i) \simeq 0.5 \, \delta_i M_{\text{\tiny H},i}
(H_i t)^{2/3},
\ee
for times larger than $t_\ita$. 

One can check that Hawking evaporation is largely subdominant with respect to accretion in affecting the PBH mass evolution. Indeed, the rate of change of the PBH mass due to Hawking evaporation is given by~\cite{MacGibbon:1991tj}
\be
\dot M_\PBH = - 0.17 f(M_\PBH) M_\PBH^{-2} M_\text{\tiny p}^4,
\ee
in terms of the function $f(M_\PBH)$ which describes the number of emitted particles and ranges from $\mathcal{O}(1)$ to $\mathcal{O}(10)$ depending on the particular PBH mass. 
For simplicity we will set it to unity, such that the solution is given by
\begin{align}
M_\PBH (t) 
& \approx M_\PBH (t_i) \llp 1 - \frac{M_\text{\tiny p}^2}{8 \pi M^2_\PBH (t_i)} (H_i t) \rrp^{1/3},
\end{align}
where we have assumed that $H_i \approx 4 \pi M_\text{\tiny p}^2/ M_\PBH (t_i)$. Since $M_\PBH (t_i) \gg M_\text{\tiny p}$, this implies that the time evolution due to Hawking evaporation is much slower than the one related to accretion. We therefore do not consider it further in our results.

The accretion flow is sufficiently spherical for the fluid to be directly accreted onto the PBH only if the tangential velocity $v_\tt$ is smaller than the Keplerian velocity in the proximity of the PBH. If not so, a disk can form, thus limiting the efficiency of the accretion rate~\cite{Ricotti:2007au}. 
The conservation of angular momentum at any radii $r$ gives
\be
v_\tt (r) r \sim \sigma_v r_\text{\tiny ta},
\ee
in terms of the typical square root of the variance of the velocity perturbation $\sigma_v(t) \simeq (3/2)^{1/3} \sigma_\text{\tiny H} (H_i t)^{1/3}$, where $\sigma_\text{\tiny H}$ is the square root of the overdensity variance at horizon crossing and we have used linear perturbation theory to determine the time dependence of the velocity perturbations. 
For a fixed initial BH mass, imposing the condition of no-disk formation translates into an upper bound on the available time for PBH mass accretion
\be
\label{nodisktime}
t \lsim \delta_i^{6/5} \sigma_\text{\tiny H}^{-9/5} G M_{\text{\tiny H},i}.
\ee
Using Eq.~\eqref{massevoaccretion}, we find that PBHs  would then accrete at most up to a maximum mass
\be
\label{maximummassacc}
M^\text{\tiny max}_\PBH \approx 0.3 \, \delta_i^{9/5} \sigma_\text{\tiny H}^{-6/5} M_{\text{\tiny H},i}.
\ee
In the case in which PBHs are formed by direct collapse and the role of spin in the collapse is negligible, that is for $\sigma_\text{\tiny H}\gsim 0.005$ \cite{Harada:2017fjm}, the PBH mass may increase by a factor $\lsim 10^2 \delta_i^{9/5}$. PBHs generated from sizeable  overdensities would therefore increase their masses by some orders of magnitude. 
Such a growth may be even larger for $\sigma_\text{\tiny H}\lsim 0.005$, when the role of the spin in the collapse is relevant, but corresponding to smaller PBH abundances.

The maximum PBH mass in Eq.~\eqref{maximummassacc} can be obtained only if the reheating time is long enough  (i.e. if the upper bound of Eq.~\eqref{nodisktime} is smaller than $t_\text{\tiny RH}$). This gives an  upper bound to the maximum PBH mass of
\be
M^\text{\tiny max}_\PBH \lesssim 0.4 \, (\delta_i\sigma_\text{\tiny H})^{3/5} M_{\text{\tiny H},\text{\tiny RH}},
\ee
in terms of the horizon mass at the end of the reheating phase
\be
M_{\text{\tiny H},\text{\tiny RH}} = \frac{1}{2 G^{3/2} T_\text{\tiny RH}^2} \simeq \lp \frac{ {\rm GeV} }{T_\text{\tiny RH}} \rp^2 M_\odot.
\ee
This implies that the maximum PBH mass can be potentially 
larger than the mass of a PBH formed at the last stage of the reheating epoch if $\delta_i\sigma_\text{\tiny H}\gsim 1$.
Furthermore, we stress that a small enough reheating temperature can make accretion last long enough to possibly delay the complete evaporation on the resulting heavier PBHs, modifying their evolution during the subsequent phases of the universe.

Finally, while a detailed study of the spin evolution due to accretion is beyond our scope, we note that for masses close to the maximum one in Eq.~\eqref{maximummassacc} it is expected that the BH spin will be close to extremality or anyway significant.

\vskip 0.3cm
\noindent
\paragraph*{\it 3.2}
Let us consider now the accretion of a PBH with mass $M_\PBH$ such that the corresponding horizon is of the order of or smaller than the Compton wavelength of the scalar particles surrounding it. 
This implies that we are resolving the quantum nature of the accreting scalar field and we cannot model the accretion flow as due to a collisionless fluid as done in the first part of this section, but rather as coming from a collisional scalar condensate.

We suppose that the PBH moves with a  speed $v$ through the scalar condensate with mass $M_h$ whose length scale is much bigger than the PBH size.  In this case the accretion rate reads~\cite{PhysRevD.14.3251,Hui:2016ltb}
%%%
\begin{align}
    \dot M_\PBH = \bigg\{ \begin{array}{lc}
         16\pi (G M_\PBH)^2\rho \quad & {\rm for}\,\,\xi\ll1, \\
         32\pi m(G M_\PBH)^3\rho/v \quad & {\rm for}\,\, \xi\gg1,
    \end{array}
\end{align}
%%%
where $\rho$ is the halo density near the PBH and $\xi=2\pi G M_\PBH m/v$. Assuming $\rho\sim 0.0044(G m^2)^3 M_h^4$ (i.e., the central density of the fundamental mode of the soliton) and $v\sim 0.3 GM_h m$ (i.e., its virial velocity), one obtains $\xi\sim 19 M_\PBH/M_h$~\cite{Hui:2016ltb}.
If $M_\PBH(t_i)\ll M_h$, then $\xi\ll1$ initially. Accounting for the evolution of the PBH mass due to accretion, one obtains that the condition $\xi\sim 1$ is reached on a timescale
%%%
\begin{equation}
    \tau_1 \sim \frac{5 M_\text{\tiny p}^{10}}{ m^6 M_h^4 M_\PBH(t_i)},
\end{equation}
%%%
corresponding to the time when the PBH has accreted approximately $5\%$ of the halo mass. After this time, the PBH would accrete the entire halo on a timescale
%%%
\begin{equation}
    \tau_2 \sim \frac{0.4 M_\text{\tiny p}^{10}}{ m^6 M_h^3 M^2_\PBH(t_i)}.
\end{equation}
%%%
These processes may occur only if the reheating phase lasts sufficiently long. By comparing it with the two timescales, one can then set an upper bound on the reheating temperature  for this dynamics to take place, that is
\begin{align}
   T_\text{\tiny RH} &\lesssim 0.4 \, \frac{m^{5/2} M_h^{2} M^{1/2}_\PBH(t_i)}{M_\text{\tiny p}^5} \sqrt{M_\text{\tiny p} m},
  \end{align}
  and
  \begin{align}
T_\text{\tiny RH} &\lesssim 1.3 \, \frac{m^{5/2} M_h^{3/2} M_\PBH(t_i)}{M_\text{\tiny p}^5} \sqrt{M_\text{\tiny p} m}.
\end{align}
In the case in which the PBH is originated from the collapse of a solitonic cloud, then $M_h\gsim M_\PBH(t_i)\gsim M_\text{\tiny p}^2/m$. Since $T_\text{\tiny RH}\simeq\sqrt{\gamma M_\text{\tiny p}m}$, with $\gamma\ll 1$, we conclude that, in this case, both processes have enough time to occur. Therefore, regardless of its initial mass, the PBH mass at reheating would be that of the surrounding halo.

\vskip 0.3cm
\noindent
\paragraph*{\it 4. Conclusions.}
\noindent
PBHs may play a key role either in the production of the so-far observed gravitational waves through mergers or in providing the dark matter of the universe. Their generation may take place during the evolution of the early universe when the latter experienced a matter-dominated phase. In this paper we have pointed out two cases in which accretion plays a crucial role during such phase.  First, we have shown that PBHs may have formed from initial seeds thanks to post-collapse accretion of the surrounding material. Secondly, we have found that PBHs, once formed, may accrete and increase their mass by several orders of magnitude. Accretion proves to be relevant not only for PBHs generated during the radiation phase and for sufficiently large masses~\cite{DeLuca:2020bjf, DeLuca:2020qqa}, but also for PBHs whose birth happens during a matter-dominated phase, possibly before the reheating of the universe. Our results are based on the assumption of spherical symmetry. It would be interesting (and pressing) to 
investigate, both analytically and numerically, the role of accretion in more realistic set-ups, as well as the role  of fragmentation for which, starting from one collapsing cloud,  gravitational cooling should lead to more than one accreting seed~\cite{1962ApJ...136..594H}.

\vspace{.5 cm}
%--------------------------------------------------------------------------------------
\paragraph*{ Acknowledgments.}
V.DL. and A.R. are supported by the Swiss National Science Foundation 
(SNSF), project {\sl The Non-Gaussian Universe and Cosmological Symmetries}, project number: 200020-178787.
P.P. and G.F. acknowledge financial support provided under the European Union's H2020 ERC, Starting 
Grant agreement no.~DarkGRA--757480, and under the MIUR PRIN and FARE programmes (GW-NEXT, CUP:~B84I20000100001), and support from the Amaldi Research Center funded by the MIUR program ``Dipartimento di Eccellenza" (CUP:~B81I18001170001).

\bibliography{draft}

\end{document}